%\\
%Title: Dual Descriptions of Supersymmetry Breaking
%Authors: Kenneth Intriligator and Scott Thomas
%Comments: 16 pages, harvmac
%Report-no: SLAC-PUB-7143
%\\
%
%\\

\input harvmac
%\draftmode
\noblackbox
\def\np#1#2#3{Nucl. Phys. B {#1} (#2) #3}
\def\pl#1#2#3{Phys. 
Lett. B {#1} (#2) #3}

\def\prl#1#2#3{Phys. Rev. Lett. {#1} (#2) #3}
\def\physrev#1#2#3{Phys. Rev. D {#1} (#2) #3}

\def\prep#1#2#3{Phys. Rep. {#1} (#2) #3}

\def\tilde{\widetilde}

%  draw box of size #1pt and line thickness #2pt

\def\drawbox#1#2{\hrule height#2pt 
        \hbox{\vrule width#2pt height#1pt \kern#1pt 
               \vrule width#2pt}
               \hrule height#2pt}

% Young tableaux

\def\Fund#1#2{\vcenter{\vbox{\drawbox{#1}{#2}}}}
\def\Asym#1#2{\vcenter{\vbox{\drawbox{#1}{#2}
              \kern-#2pt       % line up boxes
              \drawbox{#1}{#2}}}}
 
\def\fund{\Fund{6.5}{0.4}}
\def\asym{\Asym{6.5}{0.4}}

\baselineskip=12pt
\hfill{SLAC-PUB-7143} 

\hfill{IASSNS-HEP-95/115} 

\hfill{hep-th/9608046}

\Title{}
{\vbox{\centerline{Dual Descriptions of Supersymmetry Breaking}}} 

\bigskip
\centerline{\tenbf Kenneth Intriligator}
\vglue .1cm
\centerline{Institute for Advanced Study}
\centerline{Princeton, NJ 08540}

\vglue .2in 

\centerline{\tenbf Scott Thomas}
\vglue .1cm
\centerline{Stanford Linear Accelerator Center}
\centerline{Stanford University}
\centerline{Stanford, CA 94309}

\bigskip

\noindent

Dynamical supersymmetry breaking is considered in models which admit
descriptions in terms of electric, confined, or magnetic degrees of
freedom in various limits.  In this way, a variety of seemingly
different theories which break supersymmetry are actually
inter-related by confinement or duality.  Specific examples are given
in which there are two dual descriptions of the supersymmetry breaking
ground state.

\Date{}

\lref\review{K. Intriligator and N. Seiberg, hep-th/9509066.}

\lref\itquantum{K. Intriligator and S. Thomas, SLAC-PUB-7041,
hep-th/9603158, to appear in Nucl. Phys. B.}

\lref\intpou{K. Intriligator and P. Pouliot, hep-th/9505006, 
\pl{353}{1995}{471}.}

\lref\adssb{I. Affleck, M. Dine, and N. Seiberg, \pl{137}{1984}{187}}
\lref\adscalc{I. Affleck, M. Dine, and N. Seiberg,
\prl{52}{1984}{1677}.}

\lref\ven{Y. Meurice and G. Veneziano, \pl{141}{1984}{69}.}

\lref\ads{I. Affleck, M. Dine, and N. Seiberg, \np{241}{1984}{493}.}
\lref\adstt{I. Affleck, M. Dine, and N. Seiberg, \np{256}{1985}{557}.}

\lref\amati{D. Amati, K. Konishi, Y. Meurice, G. Rossi, and 
G. Veneziano, \prep{162}{1988}{169} and references therein.}

\lref\ils{K. Intriligator, R.G. Leigh and N. Seiberg, hep-th/9403198,
\physrev{50}{1994}{1092}.}%

\lref\sem{N. Seiberg, hep-th/9411149, \np{435}{1995}{129}.}%

\lref\iss{K. Intriligator, N. Seiberg, and S. Shenker,
hep-ph/9410203,
\pl{342}{1995}{152}.}

\lref\brp{J. Bagger, L. Randall, and E. Poppitz, hep-ph/9405345,
\np{426}{1994}{3}.}

\lref\nati{N. Seiberg, hep-th/9402044, \physrev{49}{1994}{6857}.}%

\lref\nonren{N. Seiberg, hep-ph/9309335, \pl{318}{1993}{469}.}

\lref\swi{N. Seiberg and E. Witten, hep-th/9407087,
\np{426}{1994}{19}.}% 
 
\lref\sn{A. Nelson and N. Seiberg, hep-ph/9309299, \np{426}{1994}{46}.}

\lref\hitoshi{H. Murayama, hep-ph/9505082, \pl{355}{1995}{187}.}

\lref\pt{E. Poppitz and S. Trivedi, hep-th/9507169,
\pl{365}{125}{1996}.}

\lref\panti{P. Pouliot, hep-th/9510148, \np{367}{151}{1996}.}

\lref\bkn{M. Dine and D. Macintire, hep-ph/9205227, 
\physrev{46}{2594}{1992};
T. Banks, D. Kaplan, and A. Nelson, hep-ph/9308292,
\physrev{49}{1994}{779}.}

\lref\smallmu{G. Farrar, RU-95-25, hep-ph/9508291;
J. Feng, N. Polonsky, and S. Thomas,
%SLAC-PUB-7050, 
hep-ph/9511324, \pl{370}{95}{1996}.}

\lref\tools{M. Dine, A. Nelson, Y. Nir, and Y. Shirman,
%SCIPP 95-32, 
hep-ph/9507378, \physrev{53}{1996}{2658}.}

\lref\visible{M. Dine and A. Nelson, hep-ph/9303230, 
\physrev{48}{1993}{1277};
M. Dine, A. Nelson, and Y. Shirman, hep-ph/9408384,
\physrev{51}{1995}{1362}.}

\lref\index{E. Witten, \np{202}{1982}{253}.} 

\lref\poustr{P. Pouliot and M.J. Strassler, RU-95-78, hep-th/9602031.}

\lref\talk{
K. Intriligator and S. Thomas, talk presented at {\it Unification:
{}From the Weak Scale to Planck Scale},
Institute for Theoretical
Physics, Santa Barbara, CA, Oct. 23, 1995;
Massachusetts Institute of Technology, Oct. 1995.}

\lref\popshadtriv{E. Poppitz, Y. Shadmi, S.P. Trivedi, hep-th/9605113.}
\lref\strass{M.J. Strassler, hep-th/9510342.}

\lref\noideas{C. Csaki, L. Randall, and W. Skiba, MIT-CTP-2532,
hep-th/9605108;
C. Csaki, L. Randall, W. Skiba, and R. Leigh, MIT-CTP-2543,
hep-th/9607021.}

\baselineskip=16pt

\newsec{Introduction}

Recent advances have shown that supersymmetric gauge theories can
often appear in the infrared as other theories, with different gauge
groups and matter content \sem.  This is the phenomenon of
electric-magnetic duality generalized to non-Abelian theories.  One
possibility is that two or more gauge theories, which differ in the
ultraviolet, can flow in the infrared to the same interacting fixed
point of the renormalization group.  The fixed point theory is in an
interacting non-Abelian Coulomb phase which can be described in terms
of any of the ``dual'' ultraviolet theories.  Another possibility is
that an ultraviolet theory can flow to another gauge theory which is
not asymptotically free.  The infrared theory is in a free magnetic
phase, with weakly coupled magnetic quarks and gluons.  Another
possibility is that the ultraviolet gauge theory flows to a
Wess-Zumino model without gauge interactions \nati.  The infrared
theory is described in terms of confined fields with tree level
interactions.
%Because duality and confinement involves the 
%behavior of strongly coupled gauge
%theories, it has, at present, only been proposed in the more tractable
%context of supersymmetric gauge theories.  
A review of recent work in supersymmetric gauge theories and a list of
references is given in \review.
%It is quite possible that similar phenomena also occur for
%non-supersymmetric theories.
In this paper we study chiral gauge theories in which dynamical
supersymmetry breaking admits descriptions in terms of electric,
magnetic, or confined degrees of freedom in various limits.  The
mechanism by which supersymmetry is broken in these theories is 
a dynamically generated superpotential. 
%or confinement at the origin of
%a classical moduli space.

Non-perturbative gauge dynamics can spontaneously break supersymmetry,
a phenomenon which can be useful for constructing models with natural
hierarchies of scales.  It is of interest to see how dynamical
supersymmetry breaking is compatible with duality.
%In particular, it is possible that duality can be extend
%ed, in this way,
%to non-supersymmetric theories.  In the present letter, we will
%discuss some simple examples illustrating the compatibility of duality
%and dynamical supersymmetry breaking.  
In one sense, the fact that there are dual descriptions of
supersymmetry breaking is actually standard, though perhaps somewhat
trivial.  Any theory which breaks supersymmetry dynamically with a
mass gap is dual in the far infrared to any trivial theory with the
same (discrete) vacuum structure.  Alternately, if there are gapless
excitations in the ground state, such as a $U(1)_R$ axion or massless
fermions required for anomaly cancelation, the theory is dual in the
far infrared to a non-supersymmetric non-linear sigma model.  Here we
will explore slightly less trivial notions of dual and confined
descriptions, relating supersymmetry breaking in gauge theories to
that in other gauge theories.  In order to make use of supersymmetric
dualities and confinement we consider models with (at least) two well
separated dynamical scales, $\Lambda_H \gg \Lambda_L$, with
supersymmetry broken at or below the scale $\Lambda_L$.  In this way
there is region of momenta below $\Lambda_H$, but above the
supersymmetry breaking scale, in which supersymmetry is realized
linearly.  The effective action is manifestly supersymmetric in this
region, and the powerful constraints of supersymmetry may be
implemented.

Below the scale $\Lambda_H$, the theories considered here flow
towards an interacting fixed point, free magnetic, or 
confined phase. 
At the scale $\Lambda_L$ additional strong dynamics break 
supersymmetry in the effective low energy theory.
In the case of a non-Abelian Coulomb phase, the theory
never reaches the fixed point at the origin of moduli space
since supersymmetry is broken in the ground state. 
Duality is exact only at the fixed point, but should apply
to the light degrees of freedom in the neighborhood of the fixed
point.
There can therefore be simultaneous dual descriptions of the 
supersymmetry breaking in the ground state. 
In the case of a free magnetic or confined phase there is only
one weakly coupled description of the theory below the scale 
$\Lambda_H$. 
The magnetic or confined description can however be
continuously connected to an electric description 
of the supersymmetry breaking by 
adjusting the parameters of the theory so that $\Lambda_H \rightarrow 0$
holding $\Lambda_L$ fixed.

Applications of duality and confinement to supersymmetry breaking
are useful for a number of
reasons \talk.  
As a function of the ultraviolet parameters of a theory, the
relevant degrees of freedom in the ground state are in some instances
confined or magnetic fields with a different gauge group, rather than
the underlying electric fields and ultraviolet gauge group.  Duality
or confinement is therefore required in order to give a proper
description of the supersymmetry breaking in these circumstances.  In
addition, since duality or confinement in general gives a different
low energy description of the supersymmetry breaking, it acts as a
generator for other models of supersymmetry breaking in which the
magnetic or confined fields are re-interpreted as electric fields in
the ultraviolet.  Finally, it is possible that theories which break
supersymmetry by the ``classic'' mechanism of a dynamically generated
superpotential over a classical moduli space are related by duality or
confinement to models which break supersymmetry by another mechanism.
This technique was employed in Ref. \itquantum\ to illustrate a model
which breaks supersymmetry in one limit by a dynamically generated
superpotential in one gauge group, and in another limit by the quantum
deformation of the moduli space due to another gauge group.  
%Here we
%extend this technique to illustrate a model which breaks supersymmetry
%in one limit by a dynamically generated superpotential, and in another
%limit by confinement on a classical moduli space.

In the next section we illustrate confinement and duality in a class
of simple renormalizable chiral models which break supersymmetry 
based on the gauge group $SU(N) \times SU(2)$. 
These models are related by confinement or duality to 
the $SU(N)$ models of Affleck, Dine, and Seiberg \adstt.
Some of
the models are near a non-Abelian Coulomb phase 
and admit simultaneous dual descriptions of the supersymmetry breaking
ground state.
Others are in a free magnetic, free electric, or confined
phase and only admit one weakly coupled description of the
ground state.
For these models it is possible to check the scaling of 
physical quantities such as the vacuum energy, and verify
explicitly that the weakly coupled electric and magnetic
or confined descriptions do not have overlapping regions of 
applicability. 
All these models break supersymmetry by a dynamically
generated superpotential in the electric, confined, and magnetic
descriptions, and relate generalizations of known models.  
%In section
%three we present a model with three gauge groups which in some limits
%breaks supersymmetry by a dynamically generated superpotential, and in
%another limit by confinement at the origin of a classical moduli
%space.  
In section four we discuss generalizations, and comment on
connections between models of supersymmetry breaking.  
In an Appendix
we outline non-renormalizable chiral models
based on the gauge group $SU(N) \times SP(M)$ 
which generalize the $SU(N) \times SU(2)$ models. 
%In another appendix we provide some details concerning the confined
%version of the model discussed in section three.
%comment on how almost all know chiral models 
%of supersymmetry breaking can be connected. 

%During completion of this work, several works appeared which explored
%somewhat related models \refs{\pt, \panti, \strass, \popshadtriv}.

\newsec{Supersymmetry Breaking by a Dynamical Superpotential}

Perhaps the simplest renormalizable chiral model of dynamical
supersymmetry breaking with two gauge groups is the $SU(3) \times
SU(2)$ model of Affleck, Dine, and Seiberg \adstt.  In Ref. \itquantum\
we showed that in a limit where the $SU(2)$ dynamics dominates, this
model breaks supersymmetry by the quantum deformation of the classical
moduli space.  In this section we consider $SU(N) \times SU(2)$
generalizations of this model, and show that in a limit where the
$SU(2)$ dynamics dominates, the theories are either confined, near a
non-Abelian Coulomb phase, or free magnetic phase, with gauge group
$SU(N) \times SP({1 \over 2}(N-5))$ and different matter
representations.  In all these descriptions, supersymmetry is broken
by the dynamically generated $SU(N)$ superpotential which lifts the
moduli space.

The matter content of the models is
\eqn\suspcfields{
\matrix{  
\quad     & SU(N) \times SU(2) & \quad \cr
          &               &               \cr
P         & (\fund,\fund)       & \quad         \cr
L         & (1,\fund)       & \quad         \cr
\overline{U}   & (\overline{\fund},1)   & \quad \cr 
\overline{D}   & (\overline{\fund},1)   & \quad \cr 
}} with $N$ odd.  Classically there is a moduli space of vacua
parameterized by the invariants $Z=P^2 \overline{U} \overline{D}$,
$X_1=PL \overline{D}$, and $X_2=PL \overline{U}$, with the gauge group
generically broken to $SU(N-2)\subset SU(N)$.  There is another gauge
invariant, $Y=P^NL$, which vanishes classically by Bose statistics of
the underlying fields.  At tree level there is a single renormalizable
coupling which can be added to the superpotential,
\eqn\wsutree{
W_{tree} = \lambda X_1. 
}
This superpotential leaves invariant non-anomalous %accidental
$U(1)_R$ and $U(1)$ flavor symmetries, and completely 
lifts the classical moduli space. 
Classically there is a supersymmetric ground state at the origin. 

Quantum mechanically, the non-perturbative $SU(N)$ dynamics
lifts the classical supersymmetric ground
state $Z=X_i=0$, and both the $U(1)_R$ and
supersymmetry are spontaneously broken. 
It follows from the $U(1)_R$ and $U(1)$ flavor symmetries 
that the $SU(2)$ dynamics do not lift the moduli space.
As discussed in the following subsections, 
even though only the $SU(N)$ dynamics lifts the moduli space, 
the low energy description of supersymmetry breaking 
in the ground state depends
on the relative importance of the $SU(N)$ and $SU(2)$
non-perturbative dynamics. 
This is because extra confined or magnetic degrees of
freedom associated with the $SU(2)$ can become light near the origin.
%As discussed in the next subsection,
%In a limit in which the $SU(N)$ dynamics dominated, 
%an electric description in terms of the ultraviolet fields and
%gauge groups is the relevant one. 
%However, in a limit in which the 

\subsec{Electric Description}

If the $SU(2)$ is weakly gauged in the ground state, any concomitant
non-perturbative effects may be ignored, and the $SU(2)$ may be
treated classically.  The exact superpotential over the classical
moduli space is then given by
\eqn\wsuspc{
W = (N-2) \left( {\Lambda_{N}^{3N-2} \over Z} \right)^{1/(N-2)} +
\lambda X_1.  }  
The dynamical part of this superpotential is due to
gaugino condensation in the unbroken $SU(N-2) \subset SU(N)$.  
For $\lambda \ll 1$,
the vacuum expectation values in the ground state are close to the
classical moduli space, and large compared to $\Lambda_{N}$.  In this
weak coupling limit, the relevant degrees of freedom in the ground
state are just the classical moduli, which parameterize the projection
of the elementary electric fields onto the classical moduli space.
Parametrically, for $\lambda \ll 1$, the field expectation values and
vacuum energy scale as $\phi \sim \lambda^{-(N-2)/(3N-2)} \Lambda_{N}$
and $V \sim | \lambda^{2(N+2)/(3N-2)} \Lambda_{N}^4|$.  In order for
this approximation to be valid, the non-perturbative $SU(2)$ dynamics
must be unimportant at the scale of the expectation values.  For $N
\leq 9$, the $SU(2)$ is asymptotically free and the requirement of
weak coupling amounts to $\Lambda_{N} \gg \lambda^{(N-2)/(3N-2)}
\Lambda_2$, with $\Lambda _2$ the scale of the $SU(2)$.  
For $N \geq 11$, the $SU(2)$ is not asymptotically free, and weak
coupling in the ground state requires $\Lambda_{N} \ll
\lambda^{(N-2)/(3N-2)} \Lambda_2$.  In these limits, the $SU(2)$ acts
as a spectator in the non-perturbative $SU(N)$ dynamics which breaks
supersymmetry.  Its only role is to provide a classical gauge
potential which lifts certain directions in field space.  In the
following subsections we consider the limits in which the $SU(2)$
dynamics is important.

\subsec{Confined Description}

If the $SU(2)$ is strongly coupled in the ground state, its
non-perturbative dynamics can not be ignored.  For $N=5$, in the limit
$\Lambda_2 \gg \Lambda_5$, the $SU(5)$ is weakly gauged at the scale
$\Lambda_2$ and may be treated as a weakly gauged subgroup of an
$SU(6)_F$ flavor symmetry under which $P$ and $L$ are not
distinguished.  
The $SU(2)$ theory therefore has three flavors (six
$\fund$) and confines as in Ref. \nati.  
The resulting confined theory is
$SU(5)$ with matter given by 
$\widehat{F}=F / \Lambda_2 = PL / \Lambda_2$ $\in$ $\fund$ of
$SU(5)$, and 
$\widehat{A}= A / \Lambda_2 = P^2/\Lambda_2$ $\in$ $\asym$ 
of $SU(5)$
(throughout hatted fields represent canonically normalized
confined degrees of freedom).  
%with the additional confining superpotential $W_{conf}=-A^2 F$.  
For 
expectation values much less than $\Lambda_2$ these fields, along with
$\bar{U}$ and $\bar{D}$, make up the canonically normalized degrees of
freedom, as evidenced by the t' Hooft anomaly matching conditions
\nati.  
%Below the scale $\Lambda_2$, near the origin of the moduli
%space of vacua, the matter content of the low energy $SU(5)$ theory is
%different from that in the underlying high energy theory.  
In the absence of the electric tree level superpotential 
\wsutree, the moduli
space of the low energy confined theory is parameterized by 
$X_1=F \overline{D}$, 
$X_2=F\overline{U}$, 
$Z=A \overline{U}\overline{D}$, and 
$Y=A^2 F$, subject to the
confining superpotential $W_{conf}=-Y/\Lambda _2^3$.  This
superpotential ensures that the moduli space of the confined theory
agrees with that of the high energy electric theory \nati.  
In particular, just as in the electric
theory, the gauge group is generically broken to $SU(3) \subset SU(5)$
on the moduli space of the theory with $SU(2)$ confined.
%
%Without the
%additional tree level superpotential $W_{conf}$, this moduli space
%would have a generic unbroken $SU(2) \subset SU(5)$ gauge group.
%Notice that since the moduli space of the effective theory involves
%extra confined $SU(2)$ degrees of freedom which are not part of the
%classical electric theory, the pattern of breaking for the $SU(5)$ in
%the confined theory appears to differ from that in the electric
%description.  The vacua which remain unlifted upon
%adding $W_{conf}$ agree with those of the electric description, 
%having an unbroken $SU(3)$ gauge group.

The scale $\widehat{\Lambda}_5$ of the low energy theory
is related to $\Lambda_5$ by the matching relation 
$\widehat{\Lambda}_5^{12} = \Lambda_5^{13}/ \Lambda_2$
at the scale $\Lambda_2$.
The exact superpotential of the low energy theory is then 
\eqn\wspici{
W=2\left({\Lambda _5^{13}\Lambda _2^3\over YZ}\right)^{1/2}-{Y\over
\Lambda _2^3} + \lambda X_1 .  } The first term is generated by
gaugino condensation in an unbroken $SU(2) \subset SU(3) \subset
SU(5)$, as can be verified by using the matching condition and
substituting the invariants of the low energy theory.  
The second term is the confining superpotential, 
$W_{conf} = - \widehat{A}^2 \widehat{F}$, 
which is a Yukawa coupling in the low energy theory.  
The final term is the confined operator corresponding to the
electric tree level superpotential \wsutree, and 
gives a Dirac mass $m=\lambda \Lambda_2$ to
the pair $\widehat{F}$ and $\overline{D}$ in the effective theory.

For $\lambda \Lambda_2 \gg \widehat{\Lambda}_5$, the Dirac pair 
$\widehat{F} \overline{D}$ 
may be integrated out of the effective theory.
The remaining matter fields are $\overline{U}\in \fund$ and 
$\widehat{A}\in \asym$ of $SU(5)$.  This is the original
(non-calculable) model of dynamical supersymmetry breaking given by 
Affleck, Dine, and Seiberg \adssb.  %See also \amati.
%The $SU(5)\times SU(2)$ model is thus related to this
%model in one limit. 
This theory has a $U(1)_R$ symmetry and a classical 
supersymmetric ground state at the origin.
This is  
is presumably lifted by the $SU(5)$ non-perturbative effects,
with supersymmetry broken, as argued in \refs{\adssb, \amati}.
Although this theory does not have a weak coupling limit,
the expectation values in the ground state are presumably
$\phi \sim \widehat{\Lambda}_5$ and 
$V \sim \widehat{\Lambda}_5^4$.

For $\lambda \Lambda_2 \ll \widehat{\Lambda}_5$ the Dirac pair 
$\widehat{F} \overline{D}$ 
are light compared to the dynamical scale and may not be
integrated out.  The low energy theory then amounts to the above
Affleck, Dine, Seiberg theory with an extra flavor, which was studied
(as an electric theory) in Ref. \hitoshi.  In the present context,
because of the large confining Yukawa coupling 
$W_{conf} = - \widehat{A}^2 \widehat{F}$ in
\wspici, the $Y$ modulus may be integrated out by applying its
equation of motion,
$ Y = - ( \Lambda_5^{13} \Lambda_2^9 / Z )^{1/3}$.
Substituting this constraint into \wspici\ gives precisely 
\wsuspc\ for $N=5$. 
However, the physical interpretation of the effective
superpotential is not the same as in the electric description,
since the canonically normalized degrees of freedom
are different. 
The position of the ground state in the confined theory is then
determined by a balance in the superpotential
\wspici\ between the scale dependence of the gaugino condensate and
the Dirac mass term.  Parametrically, for $\lambda \Lambda_2 \ll
\widehat{\Lambda}_5$, the field expectation values and vacuum energy
scale as $\phi \sim ( \Lambda_5^{13} / (\lambda^3 \Lambda_2^4))^{1/9}$
and $V \sim | (\lambda^{12} (\Lambda_2 / \Lambda_5)^{10})^{1/9}
\Lambda_5^4|$.  In order for this weakly coupled confined description
to be the relevant one, the expectation values in the ground state
must be much smaller than the confinement scale $\Lambda_2$.  This
requires $ \Lambda_5 \ll \lambda^{3/13} \Lambda_2$, which is the
opposite limit for applicability of the weak coupling electric
description discussed in the previous subsection.

\subsec{Dual Descriptions}

For $N=7$ and $9$, in the limit $\Lambda_2 \gg \Lambda_{N}$ , the
$SU(N)$ is weakly gauged at the scale $\Lambda_2$ and may be treated
as a weakly gauged subgroup of an $SU(N+1)_F$ flavor symmetry under
which $P$ and $L$ are not distinguished.  
The $SU(2)$ theory therefore
has ${1 \over 2} (N+1)$ flavors ($N+1$ $\fund$) and flows in the
infrared towards an interacting fixed point in a non-Abelian
Coulomb phase \sem.
For $\Lambda_N \ll \lambda^{(N-2)/(3M-2)} \Lambda_2$ the
$SU(2)$ is strongly coupled in the ground state and near the fixed
point. 
In the region of the 
fixed point there are two dual descriptions
of the interacting theory, either of which may be used to describe the
low energy theory \sem.
As discussed in the introduction, duality has only been conjectured for
interacting theories precisely at a fixed point. 
However, if the dual
descriptions make sense physically, they should apply in a
neighborhood of the fixed point.
One of the strongly coupled descriptions is in terms of the
original electric fields and gauge group \suspcfields.  

The dual
magnetic description for $N=7$ and $9$ has gauge group $SU(N) \times
\widetilde{SP}({1 \over 2} (N-5))$ with matter content
\eqn\suspdual{
\matrix{  
\quad     & SU(N) \times \widetilde{SP}({1 \over 2}(N-5)) & \quad \cr
          &               &               \cr
\widehat{A}    & (\asym,1)       & \quad         \cr
\widehat{F}    & (\fund,1)  &  \quad   \cr
\widetilde{\overline{P}} & (\overline{\fund},\fund)       & \quad  \cr
\widetilde{L}   & (1,\fund)   & \quad \cr 
\overline{U}    & (\overline{\fund},1)   & \quad \cr 
\overline{D}    & (\overline{\fund},1).   &  \quad \cr
}} 
The dual descriptions are just the Affleck, Dine Seiberg
$SU(N)$ theories with an $\asym$ and $N-4$ $\overline{\fund}$
\adstt, with the maximal $SP({1 \over 2}(N-5))$ flavor symmetry
acting on the $\overline{\fund}$ promoted to a gauge symmetry,
additional matter to cancel anomalies,
and an extra flavor of $\fund$ and $\overline{\fund}$. 
The fields 
$\widehat{A}= A / \Lambda_2 = P^2 / \Lambda_2$ and 
$\widehat{F} = F / \Lambda_2 = PL / \Lambda_2$ are
confined $SU(2)$ ``mesons'' while 
$\widetilde{\overline{P}}$ and
$\widetilde{L}$ are dual ``magnetic'' quarks
(throughout tilded fields represent canonically normalized
``magnetic'' degrees of freedom).
%``mesons''
%$\widehat{F} = PL / \Lambda_2$ in the $\fund$ of $SU(7)$ and 
%$\widehat{A} =$ in the $\asym$ of $SU(7)$,
%and dual ``magnetic quarks'' $\widetilde{\overline{P}}$ in 
%the $(\overline{\fund},\fund)$ of $SU(7) \times \widetilde{SU(2)}$
%and $\widetilde{L}$ in the $\fund$ of $\widetilde{SU(2)}$.
For expectation values much less than $\Lambda_2$,
the fields \suspdual\ 
%along with 
%the electric fields $\overline{U}$ and $\overline{D}$,
represent the canonically normalized degrees of freedom in
the dual description. 
%For expectation values much less than $\Lambda_2$,
%either the electric or magnetic description may be use 
%The magnetic description for $N=9$ has the same matter representations
%with gauge group
%$SU(N) \times \widetilde{SP(2)}$.
In the absence of the electric tree level superpotential 
\wsutree, the moduli space of the dual theories are parameterized
by 
$ Z = A \overline{U} \overline{D}$,
$ X_1 = F \overline{D}$, 
$ X_2 = F \overline{U}$,
$Y = A^{(N-1)/2} F$,
$V = A \widetilde{\overline{P}}  \widetilde{\overline{P}}$,
and 
$ R = F  \widetilde{\overline{P}} \widetilde{L}$
subject to the dual tree level superpotential
\eqn\dualtree{
W_{\widetilde{tree}} = { 1 \over \Lambda_2} 
 \left( V + R \right) .
}
This superpotential, along with the non-perturbative
dual gauge dynamics discussed below, ensure that the 
moduli space of the dual theory coincides with 
the classical moduli space of the electric theory \sem.
Just as in the electric theory, the gauge group is 
generically broken to $SU(N-2) \subset SU(N)$ 
on the dual moduli space. 
It should be noted that 
because of $D$-term constraints, the dual quarks
can not gain an expectation value on the moduli 
space.

The exact superpotential over the
dual moduli space is given by 
%For $\lambda \ll 1$, the expectation values of the ``mesons'' $A$ and
%$F$ are large compared to $\Lambda_N$.  The first two terms in
%\dualtree\ therefore lead to large masses for the dual ``quarks''
%$\overline{P}$ and $\widetilde{L}$.  Below the scale of the
%expectation values (given below) the dual quarks may be integrated
%out.
\eqn\suspdualw{
W = 2 \left( { \Lambda_N^{3N-2} \Lambda _2^3
\over YZ 
 V^{(N-5)/2}}\right)^{1/2} - { N-3 \over 2}
 \left( { Y \over \Lambda_2^{(11-N)/2} } \right)^{2/(N-3)} +
W_{\widetilde{tree}}   + \lambda X_1,
 } 
%where $Y=P^NL=\Lambda _2^{(N+1)/2}A^{(N-1)/2}F$,
%$Z=P^2\overline U\overline D=\Lambda _2A\overline U\overline D$, 
%and $V=A\overline P\overline P$.
The first term arises from gaugino
condensation in an unbroken $SU(2) \subset SU(N-2) \subset SU(N)$.  
The terms $W_{\widetilde{tree}} = \widehat{A} 
  \widetilde{\overline{P}}  \widetilde{\overline{P}} + 
  \widehat{F}  \widetilde{\overline{P}}  \widetilde{L}$
are Yukawa couplings in the dual theory. 
For $Y \neq 0$ the dual quarks gain a mass from these terms.
Gaugino condensation in the dual $\widetilde{SP}({1 \over 2}(N-5))$
then gives rise to the second term. 
The final term is the magnetic operator corresponding to the
electric tree level superpotential \wsutree, and gives
a Dirac mass $m = \lambda \Lambda_2$ to the pair 
$\widetilde{F}$ and $\overline{D}$. 

For $\lambda \Lambda_2 \gg \widehat{\Lambda}_5$, the Dirac pair
$\widehat{F} \overline{D}$ may be integrated out of the dual 
description. 
The effective theory is then the Affleck, Dine, Seiberg $SU(N)$ model
with a gauged flavor symmetry, subject to the tree level 
superpotential 
$W_{\widetilde{tree}} =   \widehat{A}
   \widetilde{\overline{P}}  \widetilde{\overline{P}}$.
Since the Yukawa coupling and dual gauge coupling are large,
the dual theory does not have a weak coupling limit.
The vacuum energy presumably scales as 
$V \sim \widehat{\Lambda}_5^4$. 

For $\lambda \Lambda_2 \ll \widehat{\Lambda}_5$, the Dirac pair 
$\widehat{F} \overline{D}$ are light compared to the
dynamical scale and can not be integrated out. 
Because of the large dual tree level Yukawa coupling and strong
dual gaugino condensation, the $Y$ and $V$ moduli may be
integrated out.
The effective superpotential in this limit is precisely 
\wsuspc.
Again, the physical interpretation of \wsuspc\ is not
the same as in the electric description since the relevant
degrees of freedom are different. 
The strong $\widetilde{SP}({1 \over 2}(N-5))$ gauge dynamics does
not allow a quantitative estimate of the physical quantities
in the ground state. 
In this limit the strong coupling would appear as large corrections
to the Kahler potential for $Z, X_i$, and $R$.

\subsec{Free Magnetic Description of Another Theory}

For $N \geq 11$ the $SU(2)$ is not asymptotically free.  In this case
the theory becomes strongly coupled at short distances and the degrees
of freedom \suspcfields\ can not be the relevant ones in the far
ultraviolet.  The question then naturally arises as to whether the
theory \suspcfields\ for $N \geq 11$ can be interpreted as an
effective weakly coupled magnetic description of another ``electric''
theory which {\it is} asymptotically free.  A class of electric
theories which are asymptotically free and have \suspcfields\ as a
free magnetic phase are
\eqn\suspmagnetic{
\matrix{  
\quad     & SU(N) \times {SP}({1 \over 2}(N-5)) & \quad \cr
          &               &               \cr
{A}            & (\asym,1)       & \quad         \cr
{\overline{P}} & (\overline{\fund},\fund)       & \quad  \cr
{ L}   & (1,\fund)   & \quad \cr 
\overline{U}   & (\overline{\fund},1)   & \quad \cr 
}} for $N$ odd.  
%These are just the Affleck, Dine, Seiberg $SU(N)$
%theories with an $\asym$ and $N-4$ $\overline{\fund}$ \adstt, with the
%maximal $SP({1 \over 2}(N-5))$ flavor symmetry acting on the
%$\overline{\fund}$ promoted to a gauge symmetry, and additional matter
%to cancel anomalies.
These theories have the same matter content as the dual 
theories in the previous subsection with the extra 
$SU(N)$ flavor removed.

In order to write the gauge invariants of this theory 
it is useful to define 
$V_{\alpha \beta} = A \overline{P}_{\alpha} \overline{P}_{\beta}$,
and $Q_{\alpha} = A \overline{P}_{\alpha} \overline{U}$,
where $\alpha, \beta$ are $SP(M)$ indices. 
The classical moduli space is then parameterized by $V^k$
and $Q V^{k-1} L$, 
$k=1,\dots,{1 \over 2}(N-5)$, with the $SP(M)$ indices
contracted with the invariant antisymmetric matrix. 
On the moduli space the gauge group is generically broken to 
$SU(5) \subset SU(N)$,
with $\asym$ and $\overline{\fund}$ of $SU(5)$ remaining. 
At tree level there is a single renormalizable coupling
which can be added to the superpotential
\eqn\suspff{
W_{{tree}} = \lambda V .
}
This superpotential leaves invariant a non-anomalous $U(1)_R$
symmetry, and completely lifts the classical moduli space. 
Classically there is a supersymmetric ground state at the origin. 

Quantum mechanically, the non-perturbative $SU(N)$ dynamics lifts the
classical supersymmetric ground state at the origin and supersymmetry
is broken.  The low energy description of supersymmetry breaking in
the ground state depends on the relative importance of the $SU(N)$ and
$SP({1 \over 2}(N-5))$ non-perturbative dynamics.  If the
$SP(M)$ is weakly coupled in the ground state, it may be
treated classically.
%The unbroken $SU(5)$ with $\asym$ and $\overline{\fund}$
%then lifts the moduli space.
The position of the ground state is then determined by a balance
between the potential generated by the unbroken $SU(5)$ with $\asym$
and $\overline{\fund}$ and the tree level potential.  For 
$\lambda \ll 1$ the expectation values along the moduli space, and
vacuum energy scale as $\phi \sim \lambda^{-13/(4N+7)}
\Lambda_{SU}$ and $V \sim \lambda^{(4N-20)/(2N+3)}
\Lambda_{SU}^4$ \adstt.  In order for this approximation to be valid,
the $SP({1 \over 2}(N-5))$ must be weakly coupled at the scale of the
expectation values, which requires $ \lambda^{-13/(4N+7)}
\Lambda_{SU} \gg \Lambda _{SP}$.

If the $SP({1 \over 2}(N-5))$ is strongly coupled in the ground state,
its non-perturbative dynamics can not be ignored.  
For $\Lambda_{SP} \gg \Lambda_{SU}$, 
$SU(N)$ is weakly gauged at the scale
$\Lambda_{SP}$, and may be treated as a weakly gauged flavor symmetry.
The $SP({1 \over 2}(N-5))$ therefore has ${1 \over 2}(N+1)$ flavors
($N+1$ $\fund$) and for $N \geq 11$ flows in the infrared towards a
weakly coupled theory in a free magnetic phase.  
The weakly coupled
magnetic description has gauge group $SU(N) \times \widetilde{SU}(2)$ with
``mesons'' 
$\widehat{\overline{A}}=\overline{A} / \Lambda_{SP} = 
  \overline{P}^2 / \Lambda_{SP}$ $\in$
$\overline{\asym}$ of $SU(N)$ and 
$\widehat{\overline{D}} = \overline{D} / \Lambda_{SP} = 
\overline{P} L/ \Lambda_{SP}$ $\in$ $\overline{\fund}$ of
$SU(N)$, and dual ``magnetic'' quarks $\widetilde{P}$ 
$\in$ $(\fund,\fund)$ of 
$SU(N) \times \widetilde{SU}(2)$ 
and $\widetilde{L}$ $\in$ $\fund$ of $\widetilde{SU}(2)$.  For
expectation values much less than $\Lambda_{SP}$ these fields, along
with the electric fields $A$ and $\overline{U}$ make up the
canonically normalized degrees of freedom.  The matter content of this
free magnetic phase is just that of the $SU(N) \times SU(2)$ model
\suspcfields\ with an
additional flavor of $\asym$ and $\overline{\asym}$ of $SU(N)$.  The
scale $\widehat{\Lambda}_{SU}$ of the magnetic theory is related to
$\Lambda_{SU}$ by the matching condition $\widehat{\Lambda}_{SU}^{2N}
= \Lambda_{SU}^{2N+3} /
\Lambda_{SP}^3$ at the scale $\Lambda_{SP}$.

In the absence of the electric tree level superpotential \suspff,
the moduli space of the free magnetic theory is 
parameterized by 
$Z = \widetilde{P}^2 \overline{U} {\overline{D}}$,
$X_1 = \widetilde{P} \widetilde{L} {\overline{D}}$,
$X_2 = \widetilde{P} \widetilde{L} \overline{U}$,
$\overline{V} = \widehat{\overline{A}} \widetilde{P}\widetilde{P}$,
and 
$V = {\overline{A}} A$,
subject to the dual tree level superpotential
\eqn\dualmagtree{
 W_{\widetilde{tree}} = 
    { 1 \over \Lambda_{SP}} \left( \overline{V} + X_1 \right)  .
}
This superpotential, along with the non-perturbative 
$\widetilde{SU}(2)$ dynamics ensures that the moduli space of 
the free magnetic theory coincides with classical moduli space
of the electric theory. 
With the electric tree level superpotential \suspff, 
the full tree level superpotential in the magnetic theory is 
\eqn\wsuspmtree{
W_{tree} = W_{\widetilde{tree}} + \lambda V .
}
It follows from symmetries, holomorphy, and limits that there
are no additional contributions to the magnetic 
tree level superpotential.   
The final term is a Dirac mass 
$m = \lambda \Lambda_{SP}$
for the pair $A$ and $\widehat{\overline{A}}$.
 For $\lambda \Lambda_{SP} \gg \Lambda_{SU}$ the Dirac
pair is much heavier than the dynamical scale in the 
free magnetic theory and may be integrated out.
Below
the scale $\lambda \Lambda_{SU}$, the effective magnetic theory is
then given by \suspcfields\ with superpotential \wsuspc\ and matching
condition $\widehat{\Lambda}_N^{3N-2}= (\lambda
\Lambda_{SP})^{N-2} \widehat{\Lambda}_{SU}^{2N} = \lambda
^{N-2}\Lambda _{SP}^{N-5} \Lambda _{SU}^{2N+3}$.
%and $\Lambda _2=\Lambda _{SP}$.
In this limit we therefore see that the free magnetic description of
\suspmagnetic\ is precisely the $SU(N) \times SU(2)$ 
theory \suspcfields\ for $N \geq 11$. 
 Note that the 
Yukawa coupling in the effective theory, 
$W =  \widetilde{P} \widetilde{L} \widehat{{\overline{D}}}$, 
is not small, so
in this limit the $SU(N)$ is not weakly coupled in the ground state.
The vacuum energy presumably scales as $V \sim
\widehat{\Lambda}_{N}^4$.

%.... integrate out AAbar, return to N-2 model....

%and matter fields given by \suspcfields.

%The strong $SP({1 \over 2}(N-5))$ dynamics must generate
%a Yukawa coupling in the free magnetic phase given by 

\newsec{Generalizations}

%Subtleties associated with 
%A number of models have appeared which naively appeared 
%to break supersymmetry based on an analysis of the equations of 
%motion, but which actually contained run-away directions. 

There are many generalizations of the applications of 
duality and confinement to supersymmetry breaking introduced
here and in \talk. 
A direct generalization of the $SU(N) \times SU(2)$ models
are $SU(N) \times SP(M)$ models discussed in the Appendix. 
Generalizations to other product gauge groups with similar
matter content are straightforward.

Most known models of dynamical supersymmetry breaking
%and all those
%discussed in this paper, 
can be obtained from $SU(N)$ with $A\in
\asym$ and $N-4$ fields $\in$ $\overline{\fund}$
by reducing the $SU(N)$ to product gauge groups \tools.
This can be accomplished in a full theory with the 
addition of vector like matter to spontaneously break
the original gauge group to a product. 
The addition of vector matter can not affect supersymmetry
breaking in the full theory. 
Consider the case of adding an adjoint, $\Phi$, with 
general renormalizable superpotential
\eqn\wtreead{
W=m\Phi ^2+g\Phi ^3.
}
For large $m$, this theory has,
in addition to the original $SU(N)$ theory at $\Phi=0$,
a variety of disconnected vacua with
low energy effective theories given by
\eqn\susubfields{
\matrix{
\quad     & SU(n_1) \times SU(n_2) \times U(1)& \quad \cr
          &               &               \cr
A_1       & (\asym ,1)_{2n_2} &\quad \cr
A_2       & (1,\asym)_{-2n_1}&\quad \cr
P    & (\fund,\fund)_{n_2-n_1}       & \quad         \cr
\overline{Q}_i         & (\overline{\fund}, 1)_{-n_2}
& \quad i=1\dots n_1+n_2-4         \cr
\overline{Q}_i'   & (1,\overline{\fund})_{n_1}   & \quad i=1\dots
n_1+n_2-4,\cr }} with $N=n_1+n_2$ odd.  
Due to the massless charged
matter, the $U(1)$ has no interesting dynamics.
The classical $U(1)$ $D$-terms are however crucial in some
instances in lifting certain directions in field space
\refs{\tools, \noideas}.
Starting from \susubfields\ with $n_1=3$, $n_2=2$ results in the
$SU(3) \times SU(2) \times U(1)$ model. 
For some models it is possible to simply ignore the $U(1)$ and
remove some matter from \susubfields\ to obtain an effective 
theory which breaks supersymmetry, although this is not
gauranteed \tools. 
Starting with $n_2=2$, ignoring the $U(1)$, removing both
$A_1$ and $A_2$,
and $(n_1 + n_2 -6)$ $\overline{Q}_i$,
and $(n_1 + n_2 -5)$ $\overline{Q}_i'$  
gives the $SU(N) \times SU(2)$ models.
Ignoring the $U(1)$, removing $A_1$ and $A_2$,
and $(n_1-4)$ $\overline{Q}_i$, and 
$(n_2-4)$ $\overline{Q}_i'$, gives 
$SU(N) \times SU(M)$ models \popshadtriv.

As another possibility 
consider adding an flavor of 
$\Omega$ and $\overline {\Omega}$ $ \in$ 
$\asym$ and $\overline{\asym}$ 
with the (non-renormalizable) superpotential
\eqn\wtreead{
W=m \Omega \overline \Omega +g(\Omega \overline \Omega)^2.
}
For large $m$, this theory yields a variety of disconnected vacua with
low energy theories
\eqn\suspabfields{
\matrix{
\quad     & SU(n_0) \times SP(n_1) & \quad \cr
          &               &               \cr
A_0       & (\asym ,1) &\quad \cr
A_1       & (1,\asym )&\quad \cr
P         & (\fund,\fund)       & \quad         \cr
\overline{Q}_i         & (\overline{\fund}, 1)
& \quad i=1\dots n_0+2n_1-4         \cr
{L_i}   & (1,{\fund})   & \quad i=1\dots
n_0+2n_1-4,\cr
}}
with $N=n_0+2n_1$.
The field $A_1$ can be given a mass by 
adding a term $W = A^2\tilde \Omega^2$ to the superpotential.
These theories are similar to the ones in sections (2.3) and 
(2.4). 
Removing $A_0$ and $(n_0 -4)$ $\overline{Q}_i$, and 
$(n_0 + 2 n_1 -5)$ $L_i$ gives the $SU(N) \times SP(M)$
models of the Appendix. 
Application of confinement or duality to any models
obtained from \susubfields\ or \suspabfields\ 
is also straightforward. 

\newsec{Conclusions}

Identifying the relevant low energy degrees of freedom in models
of supersymmetry breaking is important in giving a proper
description of the ground state. 
The simplest example is for weakly coupled theories
with a single dynamical scale which break supersymmetry
by a dynamically generated superpotential. 
The ground state sits near the classical moduli space and the 
relevant degrees of freedom are the classical moduli 
fields. 
In addition, the relevant non-perturbative superpotential 
in this limit is the exact superpotential over the classical 
moduli space. 
As demonstrated here, if the ground state sits at strong coupling
near the origin of moduli space, additional light non-perturbative
degrees of freedom can be relevant to the low energy
description of supersymmetry breaking. 
In fact for theories with multiple dynamical scales it is 
quite likely that some of the low energy degrees of freedom 
are confined or magnetic. 

If the strong dynamics is confining or in a free magnetic phase
a weak coupling description
of the supersymmetry breaking
in terms of different gauge groups and matter content than the 
electric description
can be obtained in this regime.
By adjusting parameters of the model the confined or free magnetic
description can often be continously connected to 
a weakly coupled electric description. 
Supersymmetry breaking can be explicitly verified in both 
limits. 
It is important to note that since supersymmetry is broken,
there can in principle
be phase transitions as a function of the parameters
of the model.
So the existence of two weakly coupled descriptions
does not gaurantee that supersymmetry is broken for all values
of the parameters. 
However, the confinement or duality in this case 
does act as a generator for another weakly coupled
model of supersymmetry breaking. 

If the strong dynamics is in a non-Abelian Coulomb phase
near an interacting fixed point the dual descriptions
formally do not have a weak coupling limit.
In this case there is more than one interacting
description of the supersymmetry breaking. 
By adjusting the parameters of the theory to move the
ground state far enough from the fixed point, one
of the duals is more weakly coupled 
and provides the most relevant description.  

Confinement and duality in supersymmetry breaking models
can also have important phenomenological consequences
for parameters of the low energy theory.  
A confined or magnetic description can have mass
scales which appear unnatural in the low energy theory,
as for the mass of Dirac pair in the 
confined or dual descriptions of the $SU(N) \times SU(2)$ 
models. 
Alternately, a small dimensionless parameter can appear
in a renormalizable low energy theory if the underlying
electric theory is non-renormalizable.

\centerline{{\bf Acknowledgments}}

We would like to thank M. Dine, A. Nelson, and N. Seiberg
for useful discussions.  The work of K.I. was supported by the Rutgers
DOE grant DE-FG05-90ER40559, the I.A.S. NSF grant PHY-9513835, and the
W.M. Keck Foundation.  The work of S.T. was supported by the
Department of Energy under contract DE-AC03-76SF00515.  We would like
to thank the Rutgers physics department and Aspen Center for
Physics were this work was partially completed.

\appendix{A}{$SU(N)\times SP(M)$ Generalizations}

The $SU(N) \times SU(2)$ models of section 2 may be generalized
to $SU(N) \times SP(M)$ models with matter content 
\eqn\suspgfields{
\matrix{  
\quad     & SU(N) \times SP(M) & \quad \cr
          &               &               \cr
P         & (\fund,\fund)       & \quad         \cr
L         & (1,\fund)       & \quad         \cr
\overline{Q}_i   & (\overline{\fund},1)   & \quad i=1\dots 2M, \cr 
}} 
with $N$ odd.  For any $N$ and $M$, a superpotential is generated
by one (and only one) of the two gauge groups: 
for $M \leq {1 \over 2} ( N-1)$  it is
generated by $SU(N)$ dynamics and for 
$M \geq {1 \over 2}(N+1) $ it is generated by $SP(M)$ dynamics. 
The electric version of the models with $M \leq {1 \over 2} (N-1)$ 
were discussed in Ref. \tools. 
The quantum modification of the moduli space for $M = {1 \over 2}(N-1)$
and quantum removal of flat directions for $M={1 \over 2}(N+1)$ were
discussed in Ref. \itquantum.
The models with $M \leq {1 \over 2} (N-3)$ can have confining or dual
descriptions analogous to the $SU(N) \times SU(2)$ models
of section 2. 

For $M \leq {1 \over 2} (N-1)$ the classical moduli space is 
parameterized by 
$Z_{ij}=P^2\overline
Q_{[i}\overline Q_{j]}$ and $X_i=PL\overline Q_i$
with the gauge group generically broken to 
$SU(N-2M) \subset SU(N)$. 
The tree level superpotential 
\eqn\wsutreeg{
W_{tree}=\lambda X_1+\sum _{i,j>2}\gamma ^{ij}Z_{ij}.
}
completely lifts the moduli space \refs{\tools, \itquantum}.
Classically there is a supersymmetric ground state at the 
origin. 
Quantum mechanically the non-perturbative $SU(N)$ dynamics lifts
the classical supersymmetric ground state and supersymmetry 
is broken. 
As with the $SU(N) \times SU(2)$ models, the relevant description
of the supersymmetry breaking ground state depends on the
relative importance of the $SU(N)$ and $SP(M)$
non-perturbative dynamics.

In
order to have a weakly coupled regime below the scale of the
non-renormalizable operators which appear in \wsutreeg, 
we assume that $\gamma ^{ij}\ll
\Lambda _{SU}^{-1}$ and $\Lambda _{SP}^{-1}$. The $Z_{ij}$ $i,j \geq
2$ moduli are lifted only the non-renormalizable terms in \wsutreeg\
and therefore have a much smaller classical potential than the other
moduli in this limit.  As a consequence, for $\lambda \ll 1$, the
quantum mechanical ground state develops large expectation values
along these directions.  
The resulting low energy theories are precisely
the $SU(N) \times SU(2)$ models of 
section 2, with $N\rightarrow N+2-2M$, along with the light
singlets $Z_{ij}$, $i,j\geq 2$, with the superpotential
\wsutreeg.  
If the $SP(M)$ is weakly coupled at the scale of the 
expectation values, this is the relevant description of 
the ground state. 
In the following we briefly mention aspects of the limits
in which the $SP(M)$ dynamics is important.

%For $2M=N-1$, $SP(M)$ has a moduli space with a quantum deformed
%constraint \intpou; this case was analyzed in detail in
%\itquantum. 
For $2M+3\leq N<6M+5$ the 
$SP(M)$ theory is asymptotically free and can be dualized as in
\refs{\sem, \intpou} to the theory
\eqn\suspdualg{
\matrix{  
\quad     & SU(N) \times SP(\widetilde M) 
& \quad \cr
          & & \cr
A & (\asym,1)       & \quad         \cr
F & (\fund,1)  &  \quad   \cr
\overline{P} & (\overline{\fund},\fund)       & \quad  \cr
\widetilde{L}   & (1,\fund)   & \quad \cr 
\overline{Q_i}   & (\overline{\fund},1)   & \quad i=1\dots 2M,\cr 
}} with $\widetilde M\equiv \half (N-2M-3)$, and tree level
superpotential
\eqn\wdualtreeg{
W_{\widetilde{tree}}=
A\overline P\overline P+F\overline P\widetilde L+\lambda \Lambda
_{SP}F\overline Q_{1}+\sum _{i,j>2}\gamma ^{ij}
\Lambda _{SP}A\overline Q_i\overline Q_j.
}
As in section 2, $A=P^2/\Lambda _{SP}$ and $F=PL/\Lambda _{SP}$.  
Note that the non-renormalizable term in the 
tree level superpotential 
\wsutreeg\ has become
renormalizable in the dual theory.  
For $\gamma \Lambda _{SP}\ll 1$,
$A\overline Q_i\overline Q_j$ get large expectation values, leading to
\suspdual\ with $N\rightarrow N+2-2M$ as the low energy theory.

For $2M=N-3$, the $SP(\widetilde M)$ in \suspdualg\ is trivial,
revealing that $SP(M)$ confines.  
The confined theory
\suspdualg\  in this case was
discussed as an electric theory in \pt.  
The relation between
\suspgfields\ and \suspdualg\ for $2M=N-3$ was also noted in
\panti, where it was generalized to include additional $SU(N)$
fundamental flavors.  
For $2M+3<N\leq 3M+2$, the $SP(\widetilde M)$
dual in \suspdualg\ is not asymptotically free and is therefore free
in the infrared; this range does not occur for the $M=1$ theories
discussed in section 2.  
For $3M+2<N<6M+5$, both the original theory 
\suspmagnetic\ and its dual \suspdualg\ flow to an interacting fixed
point without a supersymmetric vacuum.

%The invariants are given by
%$M_i=F\overline Q_i$, $Z_{ij}=A\overline Q_i\overline Q_j$, and
%$Y=A^{(N-1)/2}F$, which are the left uneaten by the Higgsing
%$SU(N)\rightarrow SU(2)$ on the moduli space.  The superpotential for
%in this case is 

For $N\geq 6M+5$, \suspgfields\ is not asymptotically free but can be
interpreted as the low energy description of the asymptotically free
theory
\eqn\suspmagneticg{
\matrix{  
\quad     & SU(N) \times SP(\widetilde M) & \quad \cr
          & & \cr {A} & (\asym,1) & \quad \cr {\overline{P}} &
(\overline{\fund},\fund) & \quad \cr 
\widetilde {L} & (1,\fund) & \quad \cr
\overline{Q_i}   & (\overline{\fund},1)   & \quad i=1\dots 2M,\cr 
}} with $\widetilde M=\half (N-3-2M)$ and the general, renormalizable,
tree level superpotential
\eqn\suspffg{
W_{tree}=\widetilde \lambda A\overline P\overline P+\sum
_{i,j}\widetilde \gamma ^{ij}A\overline Q_i\overline Q_j.
}  For
$\widetilde \gamma\ll 1$, $A\overline Q_i\overline Q_j$ get large
expectation values, leading to \suspmagnetic\ with $N\rightarrow
N+2-2M$ as the low energy theory.  Dualizing $SP(\widetilde M)$,
\suspmagneticg\ leads to
\suspgfields\ with $\overline Q_1=\overline
P^2\widetilde L/\Lambda _{\widetilde{SP}}$ and additional fields
$A\in\asym$ and $\overline A=\overline P^2/\Lambda _{\widetilde{SP}}
\in\overline{\asym}$ of $SU(N)$, and a
superpotential
\eqn\wsuspmtreeg{W_{tree}=\overline AP^2+\overline Q_1PL+\widetilde 
\lambda \Lambda _{SP}A\overline A +
\sum _{i,j}\widetilde \gamma ^{ij}A\overline Q_i\overline Q_j.}
Integrating out the massive $A\overline A$ pair results in the theory
\suspgfields\ with the tree level superpotential \wsutreeg\ with
$\lambda =1$ and $\gamma =-\tilde \gamma (\widetilde \lambda 
\Lambda _{SP})^{-1}$.  The
non-renormalizable theories discussed in the beginning of this
appendix are thus obtained as the low energy description of
renormalizable theories.

\listrefs
\bye